\documentclass[10pt,conference]{IEEEtran}

\usepackage[numbers]{natbib} 
\usepackage{comment}
\usepackage{graphicx}
\usepackage{multicol}
\usepackage{multirow}
\usepackage{amsmath} 
\usepackage{hyperref}
\usepackage{cleveref}
\usepackage{url}
\usepackage{tcolorbox}
\usepackage{soul}
\usepackage{xspace}
\usepackage{xcolor}

\begin{document}

\title{Malicious and Unintentional Disclosure Risks in Large Language Models for Code Generation}


\author{
    \IEEEauthorblockN{Rafiqul Rabin}
    \IEEEauthorblockA{
        \textit{rafiqul.rabin@ul.org} \\
        Digital Safety Research Institute \\
        UL Research Institutes
    }
    \and
    \IEEEauthorblockN{Sean McGregor}
    \IEEEauthorblockA{
        \textit{sean.mcgregor@ul.org} \\
        Digital Safety Research Institute \\
        UL Research Institutes
    }
    \and
    \IEEEauthorblockN{Nick Judd}
    \IEEEauthorblockA{
        \textit{nick.judd@ul.org} \\
        Digital Safety Research Institute \\
        UL Research Institutes
    }
}

\maketitle

\begin{abstract}
This paper explores the risk that a large language model (LLM) trained for code generation on data mined from software repositories will generate content that discloses sensitive information included in its training data. We decompose this risk, known in the literature as ``unintended memorization,'' into two components: unintentional disclosure (where an LLM presents secrets to users without the user seeking them out) and malicious disclosure (where an LLM presents secrets to an attacker equipped with partial knowledge of the training data). We observe that while existing work mostly anticipates malicious disclosure, unintentional disclosure is also a concern. We describe methods to assess unintentional and malicious disclosure risks side-by-side across different releases of training datasets and models. We demonstrate these methods through an independent assessment of the Open Language Model (OLMo) family of models and its Dolma training datasets. Our results show, first, that changes in data source and processing are associated with substantial changes in unintended memorization risk; second, that the same set of operational changes may increase one risk while mitigating another; and, third, that the risk of disclosing sensitive information varies not only by prompt strategies or test datasets but also by the types of sensitive information. These contributions rely on data mining to enable greater privacy and security testing required for the LLM training data supply chain.
\end{abstract}


\section{Introduction}
\label{sec:introduction}

Today's large language models (LLMs) for code generation are trained on vast amounts of data, ensuring that some sensitive information, such as personally identifiable information (PII) or secret account credentials, will be included \cite{kocetkov2022stack, feng2022automated, krause2023pushed}. That complexity makes it crucial to measure the likelihood that sensitive information from training data will be revealed to users, and to track how that likelihood may change over the LLM development lifecycle. 

Previous work has introduced this risk as a consequence of a tendency for neural networks to inadvertently \emph{memorize} certain elements of training data \cite{carlini2019secret, thakkar2021federated, carlini2023quantifying, rabin2023memorization, ozdayi2023controlling, huang2024codesecret, al2024traces}. However, established methods for measuring unintended memorization risks in general quantify only the threat posed by \emph{malicious disclosure} --- that is, a scenario in which a malicious actor seeks to exploit an inadvertent model behavior to exfiltrate sensitive information \cite{niu2023codexleaks, huang2024codesecret, yang2024unveiling, al2024traces}. 

In this paper, we introduce complementary methods that also address unintentional discovery of sensitive information. This becomes a consideration for systems that will handle many millions of user prompts, because in that case even very low error rates imply a nontrivial number of disclosures in absolute terms. We call this \emph{unintentional disclosure} risk, and reason that in a deployment context, LLM-based systems may disclose sensitive information even when they are not prompted to do so. We present assessment methods and empirical results which support this reasoning.

We extend existing methods in the unintended memorization field by incorporating approaches to assess both malicious and unintentional disclosure of sensitive information in training data. We demonstrate how to explore the ways in which changes to training data composition and processing are associated with changes in risk exposure over iterated releases of training datasets and models. We focus our inquiry on the code generation task, a popular use case for LLMs supported by a rapidly growing literature \cite{chen2021humaneval, austin2021mbpp, roziere2023code}, and on models trained using open-source software repositories \cite{kocetkov2022stack, li2023starcoder, lozhkov2024starcoder2}. Our work extends existing data mining practices for privacy and security testing of LLM-based systems, using an assessment framework suitable for continuous integration. 

Applying these methods to a recently developed open-source model and training dataset, we demonstrate the sensitivity of a system's risk profile to small changes in how data is composed, curated, or processed. There is already a rich body of work --- and indeed a deep repertoire of folk knowledge among ML practitioners --- about how changes to training data processing can have remarkable effects on model performance \cite{kocetkov2022stack, li2023starcoder, lozhkov2024starcoder2}. Our contribution here is to demonstrate that in the case of unintended memorization risks, changes to data processing can decrease one category of risk while increasing another. This motivates additional measurements that operationalize different ways in which the same model property (such as unintended memorization) can present risk through different mechanisms, different types of sensitive information, or different patterns of user interaction, associated with different deployment contexts or threat vectors.

Finally, we observe the need to report the results of independent privacy and security assessments of open-source LLMs and datasets. We conducted third-party assessments of the open-source Open Language Model (``OLMo'') family of models \cite{groeneveld2024olmo} and their associated training dataset ``Dolma'' \cite{soldaini2024dolma}. Our findings about potential OLMo privacy and security risks presented an ethical obligation to alert model developers to those risks in the style of ``responsible disclosure'' \cite{mcgregor2024err, cattell2024coordinated} in the information security community. We touch on considerations raised by this case study for practitioners handling sensitive information in software repositories that are mined for the training data of LLM-based systems \cite{feng2022automated, krause2023pushed, huang2022leaking, lukas2023analyzing, basak2024challenges}.

\section{Unintended Memorization for Code}
\label{sec:approach}

Unintended memorization refers to an LLM's tendency to extract specific information from its training data and disclose it when generating responses to user prompts, even when the model has not been explicitly designed to reproduce content verbatim \cite{carlini2019secret, carlini2023quantifying, rabin2023memorization, yang2024unveiling, al2024traces}. A particular token or sequence of tokens is considered extractable, or \textit{memorized}, if it appears in the training data and can be generated by the model in response to certain prompts. This phenomenon is closely related to training data extraction \cite{carlini2021extracting, ozdayi2023controlling} and membership inference attacks on software repository datasets and code generation models \cite{zhang2024membership, al2024traces}, where users craft special prompts to manipulate the model into revealing sensitive information embedded in its training data.

While unintended memorization is a general behavior of LLMs, it is not necessarily a mechanism risking data exposure. In a deployed system, other components like refusal models could mitigate the tendency for unintended memorization \cite{liu2024refuse}. An LLM could also be trained on data curated to exclude sensitive data, and used in a context where memorization of non-sensitive data is a feature, not a bug \cite{feldman2020require}. The risk associated with memorization thus depends on context and use case. In the case of code generation, a primary concern is situations where the LLM reveals specific tokens from code snippets (e.g., API keys) or entire fragments of code (e.g., proprietary algorithms) that were present in its training data, potentially leading to security, privacy, and legal concerns \cite{huang2022leaking, lukas2023analyzing, niu2023codexleaks, staab2024beyond, yang2024unveiling}. Since most LLMs are built on a shared foundation of open-source data indexed from sources like GitHub, which is known to contain sensitive information \cite{feng2022automated, krause2023pushed}, the risk of data exposure is widespread across models.

Reasoning about such risk, we observed two plausible scenarios: A case where a malicious actor intends to exfiltrate secrets from training data, and a case where an unsuspecting user unintentionally encounters sensitive information they may not even be interested in seeing. In both cases, the pathway to harm \cite{hutiriNotMyVoice2024} flows through a user presenting a system with textual input and receiving output that is hazardous in the sense that it contains sensitive information. In the case of \emph{unintentional disclosure}, a user might be in receipt of sensitive information from the training data without any knowledge of what that data contains. This contrasts with a \emph{malicious disclosure} case, where an attacker deliberately seeks to extract sensitive information with prompts written with partial or specific knowledge of the training data. The literature on assessment of risk related to unintended memorization mostly focuses on assessments designed with knowledge of the training data, and for this reason does not fully address unintentional disclosure. For instance, an attacker might provide partial code snippets that are likely to surround sensitive information in the training data and ask the LLM to complete the remaining portions, which may potentially expose exact secret credentials in response. Additionally, an LLM's tendency to memorize data could unintentionally lead to the disclosure of secrets in responses, even when the user's query is unrelated to sensitive information. To explore this, we developed a methodology that assessed each type of unintended memorization risk separately.

Our assessments primarily focus on identifying potentially sensitive information --- email addresses, phone numbers, and secrets like API keys --- in training datasets and evaluate the risk of both unintentional and malicious disclosure. Because the assessment for malicious disclosure requires knowledge of training data, each combination of model release and training dataset requires its own assessment dataset. In what follows, we apply various prompting strategies to reproducibly develop these datasets for an arbitrary model and dataset release. 

\section{Methodology}
\label{sec:evaluation}

In this section, we outline our evaluation approach for measuring the propensity for unintended memorization by LLMs in software code generation across releases.
Our evaluation is model- and data-agnostic, making it applicable to any LLM and its training dataset by analyzing model responses and training samples for the disclosure of sensitive information. However, to demonstrate the unintended memorization effect, we primarily examined two versions of OLMo models with their corresponding Dolma training datasets for code generation tasks in our experiments. We conduct our assessments using Dyff \cite{chadda2024ai}, an open-source framework with an API suitable for a continuous integration context.

\subsection{Systems Under Test} 

\textbf{Multiple Versions of the OLMo Model}. 
We assessed two consecutive releases of the OLMo model family: OLMo-7B\footnote{\url{https://huggingface.co/allenai/OLMo-7B}}, and OLMo-7B-v1.7\footnote{\url{https://huggingface.co/allenai/OLMo-7B-v1.7}}.  We chose these models because they were trained separately on different versions of the same training dataset, with the corresponding models, data, and training details publicly available. This allows us to analyze multiple releases of training datasets and their corresponding trained models using the same methodology. Additionally, the model is open-source and available on Hugging Face, facilitating the reproducibility of our experiments. 

\textbf{Multiple Versions of the Dolma Dataset}. 
The Dolma dataset is 3 trillion tokens and 4.5 TB gzipped size collected from a diverse mix of web content, academic publications, codebases, books, and encyclopedic materials from different sources \cite{soldaini2024dolma}. 
Our analysis covers two separate releases of Dolma, out of six versions that have been released overall and that vary by size, data sources, and quality filtering and processing. Dolma v1.5-sample, here abbreviated as Dolma v1.5s, was used to train the initial OLMo-7B model. The code subset of Dolma v1.5s was derived from a near-deduplicated version of The Stack dataset \cite{kocetkov2022stack}. 
Later, Dolma v1.7 was used to train OLMo-7B-v1.7, which is an updated version of the original OLMo-7B model. The code subset of Dolma v1.7 was derived from StarCoder \cite{li2023starcoder}, which is essentially a subset of The Stack v1.2 dataset. It exclusively contains data from permissively licensed GitHub repositories, applies a stronger near-deduplication strategy, and was also filtered for personally identifiable information (PII) to remove it from the dataset.

Prior to our assessment, we hypothesized that the changes from Dolma-v1.5s to Dolma-v1.7 would reduce the amount of sensitive information included in the dataset. We also expected the rate at which systems include sensitive information in their responses to decrease as a result. There is limited exploration in the literature into how changes between different dataset releases and model versions affect the presence and disclosure of sensitive data. To this end, we conducted an assessment that involved scanning code-related excerpts from Dolma datasets to identify potentially sensitive information and preparing prompts to evaluate the likelihood that OLMo models would generate responses containing that sensitive information. 

\subsection{Inspecting Dolma Datasets for Sensitive Information} 

Our experiments focused on two primary categories of sensitive information: PII and secret keys, both of which are frequently targeted in privacy and security threats. Specifically, we inspected the presence of email addresses and phone numbers, which may be PII\footnote{\url{https://www.ibm.com/topics/pii}}. For secret keys, we conducted scans for high-confidence provider patterns\footnote{\href{https://docs.github.com/en/enterprise-cloud@latest/code-security/secret-scanning/introduction/supported-secret-scanning-patterns}{https://docs.github.com/en/.../supported-secret-scanning-patterns}} that are known to match API tokens, as reported on GitHub \cite{huang2024codesecret}. These types of sensitive information, if exposed or compromised, can lead to severe consequences, such as identity theft, unauthorized access, and other forms of exploitation.

To identify potentially sensitive information within Dolma, we started by inspecting the coding subsets of each dataset used in our experiments. We focused on file extensions corresponding to seven common programming languages: Python, C, C++, Java, C\#, JavaScript, and PHP. We then applied regular expressions to detect email addresses with popular domains (i.e., com, org, net, edu, gov, and int), US-format phone numbers (i.e., 11-digit numbers starting with 1), and high-confidence GitHub secret patterns within code snippets. We counted the number of unique matches for each pattern in each file and aggregated results by programming language. All sensitive information in the code snippets was tagged by replacing it with a uniform MASK token, resulting in obfuscated code snippets for further analysis. 
We manually inspected a random sample of the matches to confirm the presence of email addresses, phone numbers, and secret keys.

\subsection{Measuring Unintentional and Malicious Disclosure} 

Having identified a collection of email addresses, phone numbers, and secret keys in the training datasets, we then analyzed the responses of OLMo models in our experiments for any potential disclosure of this sensitive information during code generation. Our assessment focused on two key perspectives: unintentional disclosure and malicious disclosure.

\textbf{Unintentional Disclosure}. To assess this risk, we prompted the OLMo models with code generation tasks from benchmark datasets including HumanEval\cite{chen2021humaneval}, MBPP\cite{austin2021mbpp}, MATH\cite{saxton2019math}, and two of our synthetically generated private datasets: UNIT (a dataset of unit conversion tasks) and OBJECT (a dataset of tasks to calculate the properties of geometric shapes). These datasets serve as proxies for likely LLM user requests and are not included in Dolma. We then examined whether the LLM-generated code contained \textit{any} sensitive information identified within the Dolma training dataset. Here, we did not consider prior knowledge of Dolma when prompting the OLMo models, focusing solely on whether the LLM-generated code accidentally exposed sensitive information from its training dataset. This scenario illustrates the risk of users encountering secrets while expecting only programming or mathematical code snippets that would accomplish the real-world tasks approximated through these benchmarks.

\textbf{Malicious Disclosure}. To assess this risk, we created prompts by masking sensitive information within code snippets from the Dolma training dataset and asked the OLMo models to generate code based on these masked prompts. We then examined whether the LLM-generated code contained \textit{particular} sensitive information from the training dataset. This scenario simulates a situation where users with partial knowledge of the training dataset attempt to exploit the model to disclose specific secrets. If an attack produces potential sensitive information, but not the precise one associated with the input --- for instance, producing a phone number, but not the expected number --- then it is considered as a failed attack.

\textbf{Exploring Various Prompting Strategies}. Additionally, we conducted experiments using various prompting strategies to quantify unintended memorization, with a particular focus on malicious disclosure, where we designed prompts based on code snippets from the training datasets. Exploring different prompting strategies is another important factor in understanding how LLMs might leak sensitive information, depending on the various ways instructions are presented \cite{ozdayi2023controlling, huang2024codesecret}. We provided code snippets as prompts by masking the corresponding sensitive information and asked the LLMs to predict it \cite{lukas2023analyzing}. Alternatively, we used prefix or suffix statements from code snippets as prompts and instructed the models to complete the remaining portion \cite{huang2022leaking}.
In total, we applied six different types of masking, infilling, and completion prompts, labeled PS\_1 to PS\_6, which also involved altering the text and template. 

\textbf{Scoring Results}. In the code-generation use case, users are likely to send a system the same prompt multiple times. This is in part because, in tests, many systems fail to generate code that compiles or passes unit tests on the first try (referred to as pass@1), but may generate at least one passing output if given k opportunities (referred to as pass@k) \cite{chen2021humaneval}. Similarly, a malicious actor may simply opt for brute force and attempt the same attack multiple times. To account for this, we assessed the likelihood of whether OLMo models generate sensitive information from training data in batches of 1, 5, and 10 attempts. Results are presented as pass@k scores, which express how often at least one of $k$ model-generated code snippets for a given query disclosed sensitive information, such as email addresses, phone numbers, or secret keys. 


\begin{figure*}[ht]
    \centering
    \includegraphics[width=0.9\textwidth]{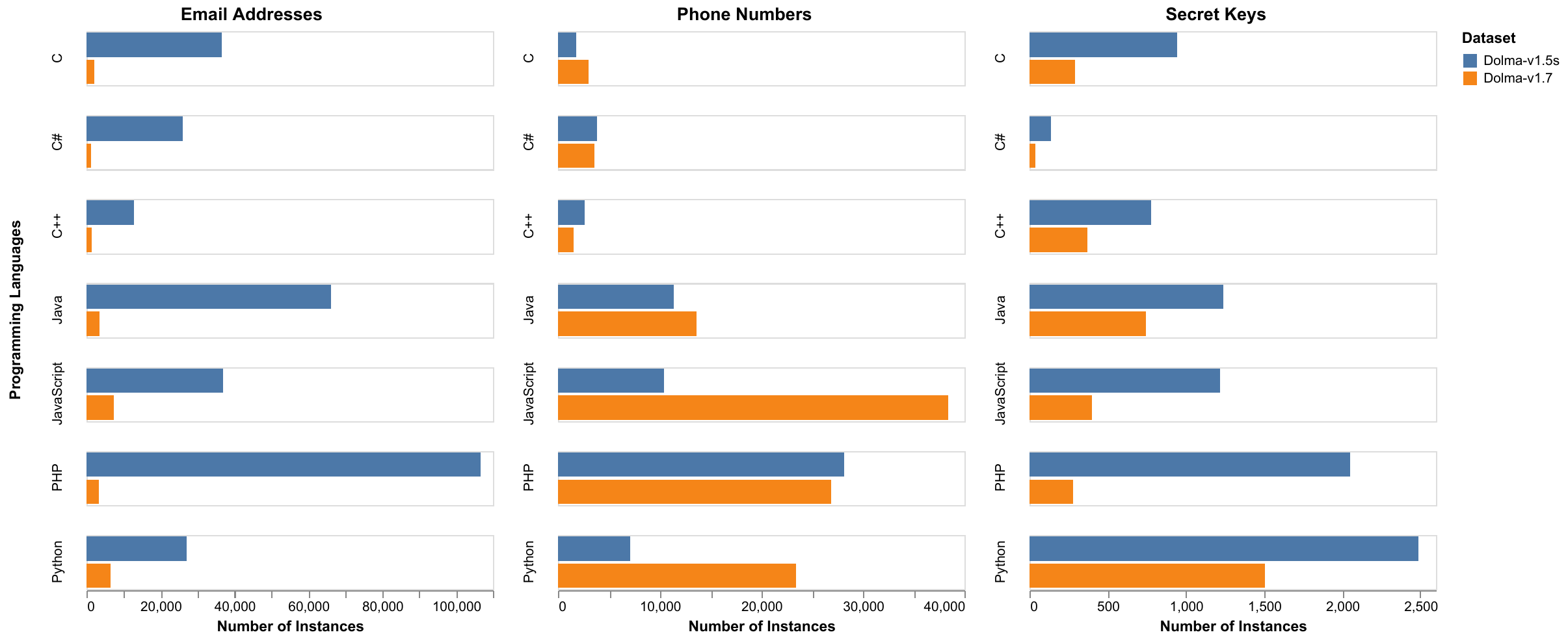}
    \caption{Count of potentially sensitive information in the Dolma training data. Results show that the count of email addresses and secret keys generally decreased between dataset releases, but the number of phone numbers increased. Analysis conducted on 89,937,427 code snippets from Dolma-v1.5s and 93,856,361 code snippets from Dolma-v1.7, aggregated across seven programming languages.}
    \label{fig:amount}
\end{figure*}

\begin{figure*}[ht]
    \centering
    \includegraphics[width=0.90\textwidth]{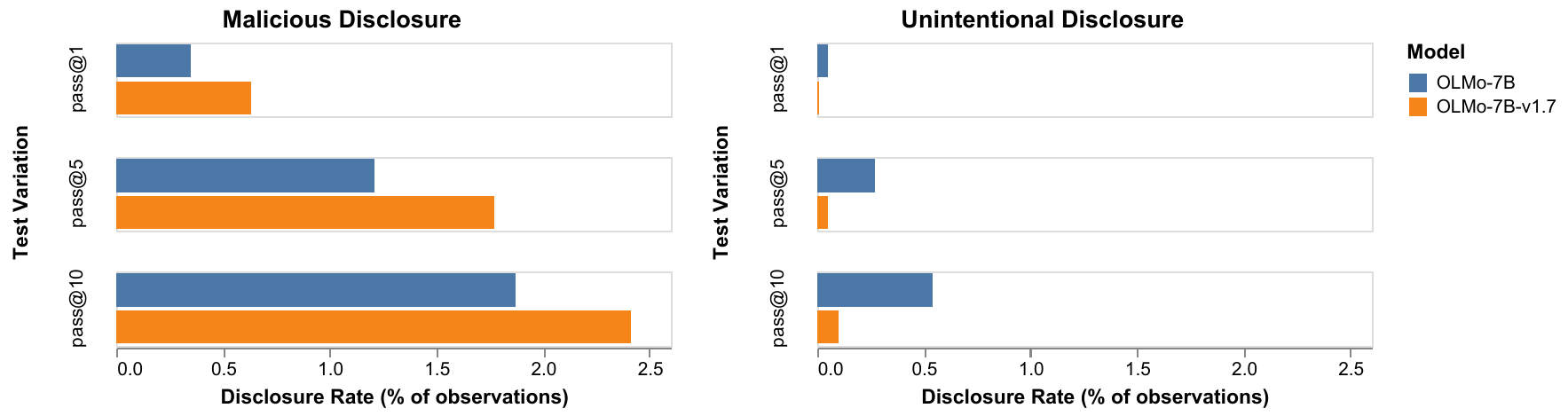}
    \caption{Model propensity to disclose sensitive information in training data. Results show that the likelihood of unintentional disclosure decreased over iterated releases of OLMo-7B, while the likelihood of malicious disclosure increased. Analysis conducted on 9,711 unique prompts for malicious disclosure and 7,973 unique prompts for unintentional disclosure for both OLMo models.}
    \label{fig:disclose}
\end{figure*}

\section{Results}
\label{sec:result}

We detail findings from our experiments with two versions of OLMo-7B models and corresponding Dolma datasets. We inspected the presence of sensitive information in training datasets, analyzed models trained on different dataset versions in the context of unintentional and malicious disclosure, and quantified the likelihood of disclosure using various prompting strategies and test datasets. We observed that the chances of data exfiltration changed between model releases --- and that at a descriptive level, data processing steps that one would expect to lead to lower chances of malicious disclosure are instead associated with increased risk.

\subsection{Prevalence of Sensitive Information in Training Dataset}

\Cref{fig:amount} compares the raw counts of potentially sensitive information, such as email addresses, phone numbers, and secret keys, included in code subsets of two separate releases of the Dolma training dataset containing source code from seven popular programming languages. 

Dolma-v1.7, which includes source code from StarCoder, contains substantially fewer likely email addresses and secret keys than Dolma-v1.5s, which derives its included source code from The Stack. However, across seven popular programming languages, the aggregate total of likely phone numbers in Dolma-v1.7 is 69\% larger than Dolma-v1.5s. 

This is interesting because in terms of dataset size measured in number of code snippets in relevant languages, the newer Dolma-v1.7 is only 4.36\% larger than its predecessor Dolma-v1.5s. Similarly, there are 16\% more code snippets containing phone numbers in Dolma-v1.7 compared to Dolma-v1.5s across seven programming languages, while Dolma-v1.7 includes 91\% and 65\% fewer code snippets containing email addresses and secret keys, respectively.\footnote{
In megabytes, the corpus containing phone numbers is marginally smaller in Dolma-v1.7 relative to Dolma-v1.5s, while the corpora containing emails and secret keys are far smaller. We take this to mean that the results are directionally the same regardless of how corpora size is measured.} 
We suspect that the increase in phone numbers is a function of how each dataset was composed and processed. While StarCoder is notable for the level of effort expended to reduce the prevalence of sensitive information it contains, the paper describing StarCoder does not mention any data preprocessing steps specifically targeting phone numbers \cite{li2023starcoder}. This implies that the StarCoder dataset was not filtered for phone numbers, despite a rigorous filtering process for other PII, such as email addresses and secret keys. The Dolma datasheet\footnote{\url{https://huggingface.co/datasets/allenai/dolma\#summary-statistics-v17}} also indicates that StarCoder was included in Dolma-v1.7 with no further processing. We speculate upon the reasons for this difference to illustrate the utility of our analysis, which highlights data processing issues for further consideration.

\subsection{Model Propensity to Disclose Sensitive Information from Training Dataset}

Moving from datasets to models, \Cref{fig:disclose} compares the likelihood of unintentional and malicious disclosure for OLMo-7B and OLMo-7B-v1.7. There are substantially greater chances of malicious disclosure for OLMo-7B-v1.7 than for OLMo-7B. Unintentional disclosure, on the other hand, is much less likely in the new release relative to the earlier one. In both cases, only a tiny fraction of prompts result in model output that is likely to contain sensitive information from training data --- up to 0.05\% of all prompts that simulate unintentional disclosure and up to 0.63\% of all prompts that represent a malicious actor, as indicated by the pass@1 score.  

In the malicious disclosure context, the newer OLMo-7B-v1.7 produced potential sensitive information in 2.41\% of cases with 10 attempts per input, representing a 29\% increase relative to the previous OLMo-7B release for pass@10 score. In other words, if an OLMo-7B-v1.7-based system is presented with 100 malicious prompts and produces output from each prompt 10 times, at least two of those 100 sets of prompts may retrieve sensitive information. In our tests, this amounts to little more than a few dozen responses out of several thousand queries. However, as previously discussed, for systems that are expected to operate at scale, processing millions or billions of distinct prompts, such a disclosure rate is concerning. 

In the unintentional disclosure context, OLMo-7B emits sensitive training data in about 0.54\% of cases with 10 attempts per input. The newer OLMo-7B-v1.7 emits sensitive information from training data in a negligible 0.1\% of cases.

\begin{figure*}[ht]
    \centering
    \includegraphics[width=0.98\textwidth]{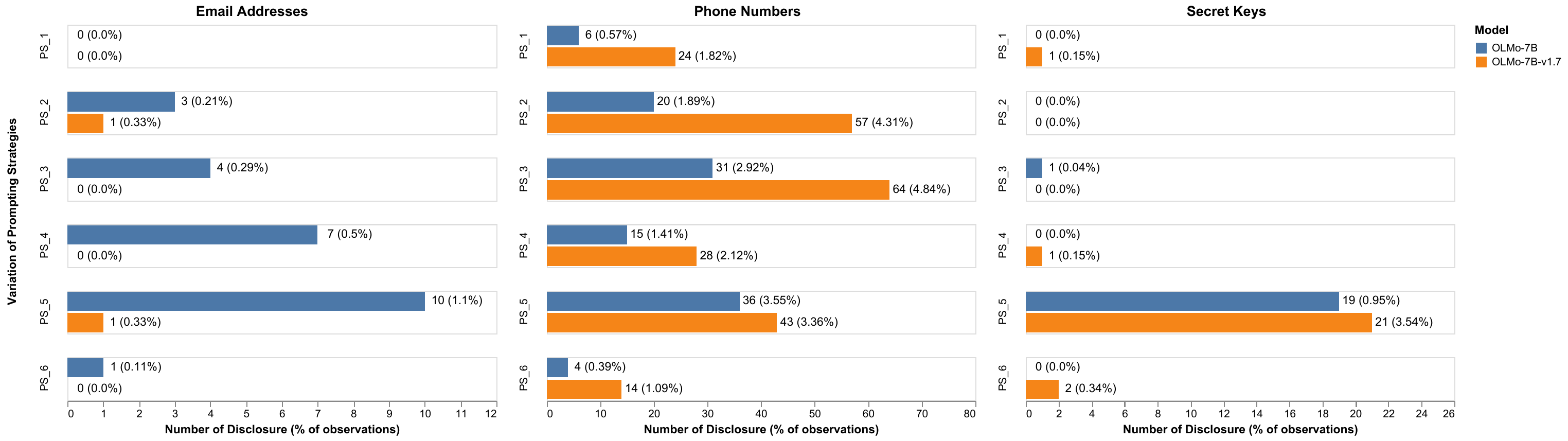}
    \caption{Malicious disclosure of sensitive information across various prompting strategies based on pass@10 results. Results show that malicious disclosures increased for phone numbers and secret keys between releases of the OLMo-7B models. For each strategy, analysis conducted on 3,917 prompts prepared from Dolma-v1.5s and 2,179 prompts prepared from Dolma-v1.7 datasets for corresponding OLMo models.}
    \label{fig:strategies}
\end{figure*}

\subsection{Malicious Disclosure by Prompting Strategy}

\Cref{fig:strategies} shows the tendency for models to comply with malicious requests for disclosure of sensitive information across different prompting strategies. The newer version of OLMo, OLMo-7B-v1.7, exposed phone numbers at a higher rate for several prompt strategies compared to the earlier version, OLMo-7B. The newer OLMo-7B-v1.7 also exposed secret keys at a higher rate with the strongest prompt strategy. However, in some cases --- such as the disclosure of email addresses by OLMo-7B-v1.7 --- the numbers are very low in absolute terms, and may be too low to reasonably extrapolate incidence rates, except when aggregated across all categories of sensitive information. 

These results indicate our method's utility in diagnosing model behaviors that result from a complex of multiple causes. For instance, our results disprove the hypothesis that simply decreasing the prevalence of a particular type of sensitive information in training data will suffice to reduce exposure risk by code-generating LLMs. If that hypothesis was correct, one would expect to see exposure rates increase in OLMo-7B-v1.7 versus OLMo-7B only for phone numbers alone, but not for secret keys.

\begin{figure*}[ht]
    \centering
    \includegraphics[width=0.85\textwidth]{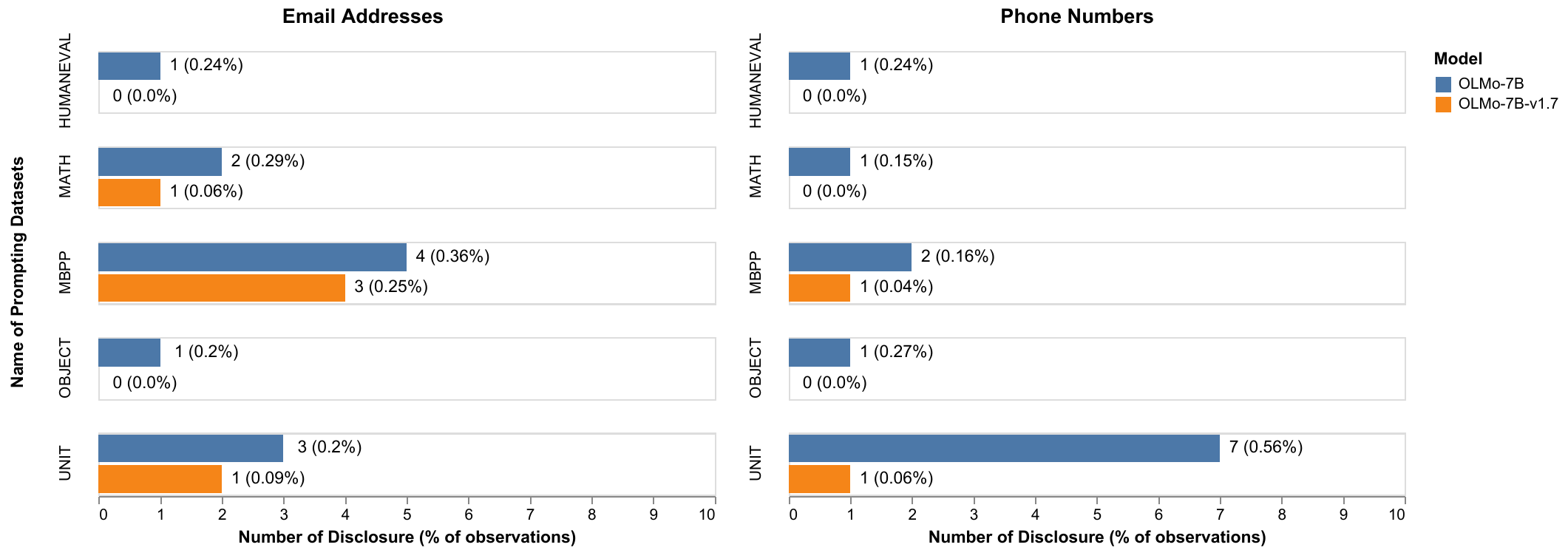}
    \caption{Unintentional disclosure of sensitive information across various test datasets based on pass@10 results. Results show that unintentional disclosure is rare for each type of sensitive information. Analysis conducted on a total of 2,897 unique prompts for both OLMo models, aggregated across test datasets.}
    \label{fig:unintentional}
\end{figure*}

\subsection{Unintentional Disclosure by Test Datasets}

\Cref{fig:unintentional} shows the tendency for models to unintentionally disclose sensitive information, disaggregated by the type of test dataset. While models may unintentionally disclose email addresses and phone numbers in some instances, they did not reveal any secret keys when generating code for programming and mathematical problems in our experiments. These results also show that unintentional disclosure is rare for each type of sensitive information. While it is hard to generalize from so few observations of a rare event, it appears that iterative model development can substantially reduce the occurrence of unintentional disclosure across datasets.

\subsection{Disclosure of Findings}

The Dolma datasets comprise material that is already public, a consideration when attempting to assess the substantive significance of the findings and weighing whether or how to disclose them to model developers and dataset compilers in advance of a more public release. However, there is another consideration: Compiling sensitive information from the broader internet into a single dataset, and developing an LLM that encodes representation of that sensitive information, changes the vectors of exposure and the likelihood that secrets will be found. For this reason, we contacted the dataset curators and model developers, the Allen Institute for Artificial Intelligence (Ai2). Ai2 maintains an email address to receive safety reports concerning their publicly available research outputs, including models and datasets. Using that email address, we contacted Ai2 to disclose our findings. Ai2 promptly acknowledged receipt, and said it would plan interventions and mitigations to address data exfiltration risks in subsequent dataset releases.

\section{Related Works}
\label{sec:related}

A substantial body of research has focused on unintended memorization across domains, including natural text \cite{carlini2021extracting, carlini2023quantifying}, code generation \cite{yang2024unveiling, al2024traces}, and others \cite{thakkar2021federated, staab2024beyond}. Many studies have also explored the disclosure of PII \cite{huang2022leaking, lukas2023analyzing}, secrets \cite{niu2023codexleaks, huang2024codesecret}, and other sensitive information \cite{yang2024unveiling, al2024traces}. However, unintentional disclosure and iterative development are not well-covered in the literature. This section provides an overview of papers relevant to our work.

\textbf{Leakage in Public Code Repositories}.
Public open-source repositories, commonly used for maintaining software code, often contain sensitive information accidentally stored by developers.
For example, \citet{feng2022automated} examined approximately 1.5 million files from about 539K GitHub repositories uploaded over a span of 75 days and found more than 142K passwords from around 64K repositories, representing about 12\% of repositories that disclosed secrets. 
GitGuardian's monitoring of public GitHub repositories also showed a two-fold increase in the number of exposed secrets in 2021 compared to 2020, totaling over six million secrets in software artifacts \cite{basak2024challenges}. 
Later, \citet{krause2023pushed} studied strategies for managing secret information in source code repositories and surveyed 109 developers about their experiences. Among the participants, 30.3\% reported encountering secret leaks, such as mistakenly pushing hard-coded secrets or \textit{.env} files to repositories. 

\textbf{Disclosure of PII and Secrets by LLMs}.
Researchers have investigated whether LLMs can disclose sensitive information by memorizing it from training datasets. 
For example, \citet{huang2022leaking} found that pre-trained GPT-Neo models do leak personal information (e.g., email addresses) due to memorization. \citet{lukas2023analyzing} next examined three types of PII leakage by GPT-2 models across three domains (law cases, healthcare, and email addresses) and extracted up to 10x more PII sequences than existing studies. 
Additionally, \citet{niu2023codexleaks} extracted sensitive personal information from the Codex model used in GitHub Copilot, including identifiable information (e.g., phone numbers, email addresses, etc.), private information (e.g., medical records, bank details, etc.), and secret information (e.g., passwords, credit card numbers, etc.). \citet{huang2024codesecret} further conducted an extensive experiment with 6 different code LLMs by exploring 18 representative secret types.

\textbf{Risks with Unintended Memorization}.
The papers exploring the leakage of sensitive information in public repositories and its disclosure by LLMs also outlined potential risks associated with such memorization. For example, \citet{yang2024unveiling} showed that memorization in code models can lead to intellectual property violations and generate security vulnerabilities. 
\citet{staab2024beyond} and \citet{niu2023codexleaks} highlighted privacy violations by pretrained LLMs on their ability to infer sensitive information from text and code, respectively. 
Additionally, \citet{katzy2024exploratory} explored the issue of license inconsistencies in LLM-generated code, highlighting legal concerns when LLMs memorize code without adhering to its licensing terms.

\section{Limitations and Directions for Future Work}
\label{sec:validity}

While this work may be of interest on matters beyond the specific goals of the paper, there are limitations and scope constraints that should be considered before applying results of our analysis to other arguments. 

First, while we grow the body of knowledge on why and how to assess the risk of unintended memorization in deployment contexts, it remains a rare event. We show the presence of memorization in both malicious and unintentional contexts. We show some variability in how likely either kind of disclosure is to occur by kind of sensitive information, prompting strategy, and/or task category. However, because each of these subgroups includes at most a few dozen successes out of thousands of trials, these results are best taken as indicative of directions for future work.

Second, while we demonstrate that unintended memorization of specific kinds of sensitive information occurs across multiple prompting strategies and across datasets for several popular programming languages in the case of a particular model family, it is unclear if exposure would be as rare --- or as frequent --- for other models. 

Third and similarly, our work is conducted in the context of a single LLM pretraining dataset. We show that our methods are valuable in part because they may aid in assessing whether data sanitization has been successful. Applying these methods to assess the relative effectiveness specific data processing steps is left for future work.

Finally, our investigation of the root cause of changes in model behavior are meant to be illustrative and exploratory, not conclusive. Our aim is to demonstrate how analyzing change in behavior across model releases can identify problems. The results presented in this work clarify the direction that a more conclusive investigation might take. 

\section{Conclusion}
\label{sec:conclusion}

This paper introduces methods to study how unintended memorization risks change during the ongoing development of LLMs for code generation. By applying the same methodology to distinct versions of the same dataset and the models they are trained on, we show that memorization risks can change substantially between one version and the next, even within the same model family. We extend the analysis of unintended memorization from malicious disclosure, where a threat actor has partial knowledge of training data, to include unintentional disclosure, where an unsuspecting user inadvertently discovers secrets. We argue that for open-source models, accidentally aggregating sensitive information from multiple repositories into a single dataset introduces new risk. When discovered, such risks should be disclosed to responsible parties. We thus expand knowledge in areas not covered by relevant literature. 

The results concerning unintentional disclosure establish that in the context of code-generating models, there is in general a nonzero chance that an unsuspecting user may encounter sensitive information --- like an email address --- in model output responding to help with some mundane programming task. The rate at which this occurs is very low, and can be brought to near zero, but it is not entirely zero. Depending on training data and deployment context, this can be a grave concern --- both for people deploying models and, possibly, people who inadvertently come into possession of data that should not be on their system.

The results also indicate that changes in data source and processing may impact the risks of sensitive data disclosure, e.g., phone numbers. We found that data pipeline changes that should be associated with more privacy-preserving model behavior was instead associated with, in some cases, increased vulnerability to a malicious user engaging in a targeted attempt to get the system to repeat specific secrets from its training data. These results show the value of testing models and datasets as part of each and every pre-release cycle for data program regressions and assess the effectiveness of mitigation strategies. This paper describes a means to conduct such tests.

\section*{Acknowledgments}

We thank our colleagues at the UL Digital Safety Research Institute and the Allen Institute for Artificial Intelligence for their valuable feedback and insights related to this paper.

\bibliographystyle{IEEEtranN} 
\bibliography{references}

\end{document}